\documentstyle[preprint,aps]{revtex}

\tightenlines

\newcommand{\vect}[1]{\mbox{\boldmath $#1$}}

\newcommand{\lsim}[1]{
\setlength{\unitlength}{12pt}
\begin{picture}(1.4,1.)
\put(.7,-0.3){\makebox(0.0,1.)[t]{$<$}}
\put(.7,-0.3){\makebox(0.0,1.)[b]{$\sim$}}
\end{picture}#1}
\begin{document}
\draft

\title{Stringent constraint on the scalar-neutrino coupling constant from 
quintessential cosmology}

\author{R. Horvat \\
   ``Rudjer Bo\v skovi\' c'' Institute, P.O.Box 1016, 10001 Zagreb,
Croatia}

\maketitle

\begin{abstract}

An extremely light ($m_{\phi} \ll 10^{-33} \;\mbox{\rm eV}$), slowly-varying
scalar field $\phi $ (quintessence) with a potential energy density as large
as $60\%$ of the critical density has been proposed as the origin of the
accelerated expansion of the Universe at present. The interaction of this
smoothly distributed component with another predominately smooth component,
the cosmic neutrino background, is studied. The slow-roll approximation for
generic $\phi $ potentials may then be used to obtain a limit on the
scalar-neutrino coupling constant, found to be many orders of magnitude more
stringent than the limits set by  observations of neutrinos from SN 1987A. In
addition, if  quintessential theory allows for a violation of the
equivalence principle in the sector of neutrinos, the current solar neutrino
data can probe such a violation at the $10^{-10}$ level.
\end{abstract}

\newpage

There are now increasing indications for a spatially flat Universe
($\Omega_0 \equiv \rho_{total}/{\rho_{crit}} = 1 $), in which a large
fraction of the present energy density 
comes from a smooth component with negative pressure that is
causing the accelerated expansion of the Universe. The simplest and at the
same time the oldest known candidate providing the necessary negative
pressure  is a non-vanishing cosmological constant. However, other 
possibilities with 
similar properties have recently been proposed,  including a dynamical, 
slowly-rolling, spatially inhomogeneous scalar field component, named
quintessence \cite{1}. The basic idea of quintessence is
that of a classically unstable field that is rolling toward its true
minimum, which is presumed to vanish. From a theoretical viewpoint,
although this does not avoid the cosmological constant problem, it still
supports a widespread belief that when the problem is properly understood,
the final answer will be zero.  From an observational viewpoint, it seems
that the best-fit models \cite{2} are those of quintessence with an
effective equation-of-state $\omega > -1$, rather than the limiting case of
the cosmological constant with $\omega = -1$.      

Beside the cosmological constant problem, there is another problem with the
rolling scalar field scenario-the initial conditions problem-
one should answer why the energy
density of the scalar field $\Omega_{\phi}$ and the matter energy density
$\Omega_m $ are of the same order of magnitude today as we know that
the energy density of the scalar field generally decreases more slowly than
that of matter. Recently, the notion of cosmological ``tracker fields''
has been introduced in certain models of quintessence \cite{3} to explain
why we live in a special era where the two densities nearly coincide.

Another set of difficulties besetting the above scenario for quintessence  
has to do with the lightness of quintessence as well as with the flatness
conditions obeyed by the potential $V(\phi )$ in any realistic model of
quintessence. In the first case \cite{4}, since  the scalar field
$\phi $ is very light (or massless) and can mediate long-range forces, it
must be subject to the constraints derived from the observational limits on
a fifth force. In the latter case \cite{5}, the flatness conditions serve 
to restrict any additional parameter (other than generic non-perturbative ones)         
in $V(\phi )$. The result of the analysis \cite{4} shows that only a moderate
suppression of a few observable interactions of quintessence with the fields
of the standard model is required, whereas the analysis \cite{5} shows that
 high-degree of fine-tuning of certain parameters in $V(\phi )$ is
required, even in the context of supersymmetry.

In the present Letter, we combine the long-range phenomenon of
quintessence with the flatness conditions for $V(\phi )$. Firstly, we
calculate the shifted mass of $\phi $ which in each point of space results 
from the interaction with the cosmic neutrino background, and then apply the
flatness conditions to restrict a scalar-neutrino coupling constant, on
which no severe constraints exist. Then we proceed by employing a mechanism
for generation of  neutrino oscillations similar to that developed in 
\cite{6}, in case an underlying quintessential theory allows for a violation 
of the equivalence principle (VEP) through a non-universal scalar-neutrino 
coupling. We find that the current solar neutrino data can then probe a VEP, 
and compare its  upper limit with the most restrictive limit for ordinary matter
\cite{7} as well as the limit on neutrinos obtained from SN 1987A \cite{8}.

We set the stage by writing down the fundamental equations governing the
above scenario for quintessence. If, for the sake of simplicity, we assume
that the total energy density has the critical value, then
\begin{equation}
V(\phi ) \sim (3 \times 10^{-3} \; \mbox{\rm eV})^{4} \;,
\label{form1}
\end{equation}
where the numerical value in (1) is the present energy density $3 M_{P}^{2}
H_{0}^{2}$. Here $M_{P} \equiv M_{Planck}/\sqrt{8 \pi } = 2.4 \times 10^{18} \;
\mbox{\rm GeV}$ is the reduced Planck scale
and $H_{0}$ is the present value of the Hubble parameter.
The effective equation of state for this component is very negative:
\begin{equation}
\omega \equiv \frac{\frac{1}{2} {\dot{\phi}}^{2} - V(\phi )}
{\frac{1}{2} {\dot{\phi}}^{2} + V(\phi )} \lsim -1/3 \;,
\label{form2}
\end{equation}
and its equation of motion is given by
\begin{equation}
\ddot{\phi} + 3H \dot{\phi} + V'(\phi ) = 0 \;,
\label{form3}
\end{equation}
where $V'(\phi )$ is the derivative of $V$ with respect to $\phi $.
In order to provide for negative pressure, quintessence should satisfy the
slow-roll condition, $3H \dot{\phi} = -V'(\phi ) $, and the necessary
conditions for the slow-roll approximation to hold are the flatness
conditions for $V(\phi )$,
\begin{equation}
M_{P} \left |V'/V \right | \ll 1 \;,
\label{form4}
\end{equation}
\begin{equation}
M^{2}_{P} \left |V''/V \right | \ll 1 \;.
\label{form5} 
\end{equation}
The application of (4) and (5) to a generic non-perturbative part in $V$ (we
may consider a potential of inverse-power, $V(\phi ) = M^{4+\alpha } \phi^{-
\alpha}$, as an
example of the tracker field, where $M$ is a parameter \cite{9})
together with the condition that $\Omega_{\phi}$ is beginning to dominate
just today, gives $\phi \sim M_{P}$ \cite{3,5}. The same for the bare mass 
term in $V(\phi )$ gives $m_{\phi} \ll 10^{-33}$. Note that for the  mass 
term in $V(\phi )$ we need consider only the second flatness condition (5), 
provided that $\phi \sim M_{P}$.

Let us now suppose that quintessence couples to neutrinos with a Yukawa
strength $g_{\nu}$ (the vacuum mass term for neutrinos is of the Dirac type), 
and consider its interaction with the background neutrinos at an effective
temperature $T_{\nu} \simeq 2 K$. We shall always treat the neutrino
component as a smooth one, since at present only $10-20\%$ of the dark
matter neutrinos are in galactic halos while the rest are distributed more
smoothly, as suggested by numerical simulations \cite{10} even for the mass
for neutrinos in the range of tens of an eV. In addition, in the light of
recent observation  of atmospheric neutrino oscillations (and hence neutrino 
mass) at Super-Kamiokande \cite{11} we consider either of two possibilities for 
neutrino  masses that are consistent with the current data: the one where the 
neutrino masses are hierarchical with the highest mass eigenvalue 
$\sim 0.1 \;\mbox{\rm eV}$, and also the case where some neutrino masses are 
nearly degenerate and larger than $\sim 0.1 \;\mbox{\rm eV}$.   

Here we would like to stress that the interaction just mentioned above would
unavoidably induce an effective mass squared $m_{\phi}^{2} + \Pi_{\phi}(0)$
in $V(\phi )$ \cite{12}, where $\Pi_{\phi}$ is the scalar self energy at finite
temperature. In this problem we take the infrared limit which is obtained by
setting the zeroth component of the external momentum to zero and taking the
limit that spatial components approach zero, i.e., $\Pi_{\phi}(k_{0}=0,
\vect{k} \rightarrow \vect{0}) \equiv \Pi_{\phi}(0)$.

For neutrino components with $m_{\nu} \ll T_{\nu}$, we find, by applying the
real-time version of Thermal Field Theory \cite{13}, at the one loop
level that
\begin{equation}
\Pi_{\phi}(0) \equiv \Pi_{\phi}^{htl}(k_{0}, \vect{k}) \simeq \frac{g_{\nu}^{2}
\; T_{\nu}^{2}}{12}\;.
\label{form6}
\end{equation}
In the opposite and more realistic case, $m_{\nu} \gg T_{\nu}$, we find
\begin{equation}
\Pi_{\phi}(0) \simeq  -0.07\; g_{\nu}^{2}\; m_{\nu}\; T_{\nu}\;,
\label{form7}                                            
\end{equation}
where the parameters in Eqs.(6) and (7) refer to the heaviest neutrino from
the background. 

Concerning Eqs. (6) and (7), a few technical remarks as well as additional
illuminations are in order. First note that, in contrast to gauge theories,
the calculation of the scalar self energy in the hard thermal loop
approximation \cite{14} shows a cancellation of the momentum-dependent terms,
leading to the simple result, Eq.(6). Furthermore, the scalar self energy
shows no imaginary part corresponding to Landau damping for all values of
the momentum and the energy. In addition, Eq.(7) represents a specific
non-equilibrium situation where massive neutrinos, which are nonrelativistic
today, still need to be assigned a distribution function relevant for
massless particles, as their total number density is fixed at about $100
\;\mbox{\rm cm}^{-3}$ in the uniform non-clustered background. 
This sort of non-equilibrium would unavoidably induce
ill-defined pinch singularities at two-loop order \cite{15} - a common feature
of out of equilibrium thermal field theories.    

With respect to the sign in Eq.(7), one should not be overmuch surprised by the
appearance of a ``thermal tachyon''. In contrast to scalar and gauge
theories, where the mass squared generated by thermal fluctuations is always
found to be positive, this is not necessarily true in a theory having
interactions which are universally attractive, signaling that a thermal
distribution just tends to collapse upon itself. Beside gravity, scalar
theories with cubic interactions in six dimensions are such another example
\cite{16}. This feature however does not show up in Eq.(6) because of free 
streaming of relativistic background particles ($m_{\nu} \ll T_{\nu}$); 
on the other hand  Eq.(7) is closely related to the 
scalar contribution to the usual Jeans mass. Finally, we have taken the number
of neutrino degrees of freedom to be equal 2 in the above equations while 
the chemical potential has been set to zero, as is probably the case for
cosmological neutrinos.     

Assuming no fine-tuned cancellations between various contributions to the
slope of $V$, we are now in position, 
by applying the flatness condition as given by Eq.(5) directly to Eqs.(6)
and (7), to set a limit on the scalar-neutrino coupling constant as
\begin{equation}
g_{\nu} \ll 10^{-28} \;\;\;\;\;\;\;(m_{\nu} \ll T_{\nu})\;,
\label{form8}
\end{equation}
\begin{equation}
g_{\nu} \ll 4 \times 10^{-30}\; \left( \frac{0.07\;\mbox{\rm eV}}
{m_{\nu}} \right )^{1/2}
\;\;\;\;\;\;\;(m_{\nu} \gg T_{\nu})\;.
\label{form9}                                        
\end{equation}

The limits (8) \footnote{It is to be noted however that even better limit in
the case where the cosmic neutrino background remains relativistic today can
be obtained by considering the neutrino mass generated by the VEV of $\phi
$, $\delta m_{\nu }=g_{\nu} \phi $. The requirement $\delta m_{\nu} \ll T_{\nu}$
then gives $g_{\nu} \ll 10^{-31}$.}
and (9) are to be compared with  the most stringent limits on 
$g_{\nu}$, that is with those  set by the
observations of neutrinos from SN 1987A. By making a claim of absence of
large scattering of supernova neutrinos from dark matter neutrinos, one
obtains, $g_{\nu} < 10^{-3}$ \cite{17}. Moreover we show that the powerful bound
as given by Eq.(9), for neutrinos having masses in the eV range, 
is even more restrictive than the corresponding limit for ordinary matter coming
from conventional solar-system gravity experiments. The present experimental
data give upper limits of order $10^{-3}$ for a possible admixture of a
scalar component to the relativistic gravitational interaction, 
$\beta_{ext}^{2} <
10^{-3}$ \cite{18}. By adjusting $g_{\nu}$ to be of gravitational origin
only, $\beta_{\nu} \equiv \sqrt{2} M_{P} (g_{\nu}/m_{\nu})$, 
one obtains from Eq.(9) that
\begin{equation}
\beta_{\nu}^{2} \ll 4 \times 10^{-6} \left( \frac{\mbox{\rm eV}}
{m_{\nu}} \right )^{3}\;.
\label{form10}       
\end{equation}

It is to be noted however that for $m_{\nu} \sim 1 \;\mbox{\rm eV}$, Eq.(10) 
represents a moderate fine-tuning in $V(\phi )$. Indeed, from a
traditional viewpoint, the expected values for the $\beta_{\nu}$ are of
order of unity since they represent interactions at the Planck scale. The
possibility to suppress such a coupling by imposing symmetries is viable
only in pseudo-Goldstone boson models of quintessence \cite {4}. Here we
give two possible solutions to the fine-tuning problem. The first solution is
in agreement with the fact that the current data favor models with a
cosmological constant over the mixed cold + hot dark matter models. In this
respect, the presence of hot dark matter is no longer necessary (and eV
neutrinos are not needed to provide this component) \cite {19}. If we set
$m_{\nu} \sim 0.04 \;\mbox{\rm eV}$, a value consistent with the
Super-Kamiokande experiment, then $\beta_{\nu} << 0.3$, and sure enough there
are ways to achieve this value without suppression by some symmetry. The
second solution is a {\sl least coupling principle} introduced originally by
Damour and Polyakov \cite {20} in string theory. It is based on a mechanism
which provides that a (universal) coupling of the scalar (the string dilaton
field in their example) with the rest of the world, 
being dependent on  the VEV of $\phi $ and hence
time dependent, has a minimum close to the present value of the $\phi $'s VEV.
Therefore, the present-day value for $\beta_{\nu}$ can naturally be much
less than unity. We have to assume however that the mechanism is also
operative for quintessence since in their paper Damour and Polyakov dealt
with the string dilaton, which, as a recent analysis \cite{21} shows, cannot
provide us with the negative equation of state, and therefore is useless for
the dynamical component of quintessence.         
      
In the rest of the paper we shall be concerned with the case where $\beta
$'s from the neutrino sector are not universal but rather species-dependent,
thereby violating the equivalence principle. Specifically this means that 
$\beta_{\nu} \rightarrow \beta_{\nu_{i}}$, where $i$ is the $i$-type neutrino. 
This is just a basic ingredient of the scenario \cite{20}, in which the
scalar may remain massless in the low-energy world and violates the
equivalence principle. By
inducing a nonzero mass squared difference for neutrinos within a medium, 
such exotic  
interactions may  produce neutrino oscillations even for degenerate-in-mass 
neutrinos \cite{6}.  

Let us now consider the case indicated in the foot-note, where a nonzero
mass squared difference for neutrinos is due to the VEP of quintessence. By
sticking with the degenerate-in-mass neutrinos, we find $\Delta m^{2}$ for
the oscillatory neutrinos 1 and 2 with the
degenerate mass $m_0$ as  $\Delta m^{2} \simeq 
2 m_0 \phi \Delta g_{\nu} $, where $\Delta g_{\nu} \equiv g_{\nu_2} -
g_{\nu_1} $. By setting $\Delta m^{2} \simeq 10^{-10}\; \mbox{\rm eV}$
as to explain the solar neutrino data via oscillation in
vacuum, one finds that
current solar neutrino data probe the quintessential scenario VEP at the
level
\begin{equation}
\Delta \beta \simeq 10^{-10} \left (\frac{\mbox{\rm eV}} {m_0} \right )^{2}
\;.
\label{form11}
\end{equation}
One finds  for neutrinos of mass $\sim 1\;\mbox{\rm eV}$
(such a model for neutrino masses where the three known
neutrinos have nearly the same mass, of about $\sim 1 \; \mbox{\rm eV}$, was
presented in Ref.\cite{22}) that the right-handed side of (11) is not
 as good as the most severe limit for ordinary matter
\cite{7}. It is however  better than the limit obtained by comparing
neutrinos
with antineutrinos from SN 1987A \cite{8}.

The limit (11) should not be confused with those obtained in 
Ref.\cite{23} as there  VEP is due to a non-universal tensor neutrino-gravity 
coupling, whereas here VEP arises due to a breakdown of universality in the 
coupling strength between the spin-0 particles and the neutrinos.  One can also
show that the above bounds remain unchanged in the case in which the vacuum
mass terms for neutrinos are of the Majorana type. \newline

{\bf Acknowledgments. } The author acknowledges the support of the Croatian
Ministry of Science and
Technology under the contract 1 -- 03 -- 068.


\begin{thebibliography}{160}

\bibitem{1} R. R. Caldwell, R. Dave and P. J. Steinhardt, Phys. Rev. Lett.
80, 1586 (1998).
\bibitem{2} L. Wang, R. R. Caldwell, J. P. Ostriker and P. J. Steinhardt,
astro-ph/9901388.
\bibitem{3} I. Zlatev, L. Wang and P. J. Steinhardt, Phys. Rev. Lett. 82,
896 (1999); P. J. Steinhardt, L. Wang and I. Zlatev, astro-ph/9812313.
\bibitem{4} S. M. Carroll, Phys. Rev. Lett. 81, 3067 (1998).
\bibitem{5} C. Kolda and D. H. Lyth, hep-ph/9811375. 
\bibitem{6} R. Horvat, Phys. Rev. D 58, 125020 (1998). 
\bibitem{7} B. R. Heckel et al, Phys. Rev. Lett. 63, 2705 (1989); E.
Adelberger et al, Phys. Rev. D 42, 3267 (1990).
\bibitem{8} M. J. Longo, Phys. Rev. Lett. 60, 173 (1988); L. M. Krauss and
S. Tremaine,  Phys. Rev. Lett. 60, 170 (1988); J. M. LoSecco, Phys. Rev. D
38, 3313 (1988); S. Pakvasa, W. Simmons and T. J. Weiler, Phys. Rev. D 39,
1761 (1989).
\bibitem{9} P. J. E. Peebles and B. Ratra, Ap. J. Lett. 325, L17 (1988).
\bibitem{10} A. L. Melott, Phys. Rev. Lett. 48, 894 (1982).
\bibitem{11} Y. Fukuda et al, Phys. Rev. Lett. 81, 1562 (1998);  Phys. Rev.
Lett. 82, 2644 (1999).
\bibitem{12} See e.g., M. E. Carrington, Phys. Rev. D 45, 2933 (1992).
\bibitem{13} A. J. Niemi and G. W. Semenoff, Nucl. Phys. B 230, 181 (1984);
R. L. Kobes and G. W. Semenoff, Nucl. Phys. B 260, 714 (1985); Nucl. Phys. B
 272, 329 (1986).
\bibitem{14} E. Braaten and R. D. Pisarski, Nucl. Phys. B 337, 569 (1990);
Nucl. Phys. B 339, 310 (1990).
\bibitem{15} T. Altherr and D. Seibert, Phys. Lett. B 333, 149 (1994); T.
Altherr, Phys. Lett. B 341, 325 (1995).
\bibitem{16} T. Grandou, M. Le Bellac and J.-L. Meunier, Z. Phys. C 43, 575
(1989).
\bibitem{17} E. W. Kolb and M. Turner, Phys. Rev. D 36, 2895 (1987); A.
Manohar, Phys. Lett. B 192, 217 (1987).
\bibitem{18}  R. D. Reasenberg et al, Astrophys. J. 234 (1979) L219.
\bibitem{19} Relic Neutrino Workshop, Sep 98, Trieste; A. Dighe, S. Pastor
and A. Yu. Smirnov, hep-ph/9812244.
\bibitem{20} T. Damour and A. M. Polyakov, Gen. Rel. Grav. 26, 1171 (1994);
Nucl. Phys. B 423, 532 (1994).
\bibitem{21} P. Binetruy,  hep-ph/9810553.
\bibitem{22} D. O. Caldwell and R. N. Mohapatra, Phys. Rev. D 48, 3259
(1993).
\bibitem{23} J. N. Bahcall, P. I. Krastev and C. N. Leung, Phys. Rev. D 52,
1770 (1995); A. Halprin, C. N. Leung and J. Pantaleone, Phys. Rev. D 53,
5365 (1996); S. W. Mansour and T. K. Kuo, hep-ph/9810510.  
\end{thebibliography}
\end{document}